# Investigating the Impacts of Recommendation Agents on Impulsive Purchase Behaviour

**Hui Zhu**
School of Management
Guangzhou University of Technology, Guangzhou, China;
School of Economics and Management
Tilburg University, The Netherlands;
Email: zhuhuistyle@126.com / H.Zhu_1@uvt.nl

**Zherui Yang**
School of Economics and Management
Tilburg University, the Netherlands;
Email: y.z_ryan@hotmail.com

**Carol XJ Ou**
Department of Management
Tilburg University, The Netherlands;
Email: carol.ou@uvt.nl

**Hongwei Liu**
School of Management
Guangzhou University of Technology, China;
Email: hwliu@gdut.edu.cn

**Robert M. Davison**
Department of Information Systems
City University of Hong Kong, Hong Kong;
Email: isrobert@cityu.edu.hk

## Abstract

The usage of recommendation agents (RAs) in the online marketplace can help consumers to locate their desired products. RAs can help consumers effectively obtain comprehensive product information and compare their candidate target products. As a result, RAs have affected consumers' shopping behaviour. In this study, we investigate the usage and the influence of RAs in the online marketplace. Based on the Stimulus-Organism-Response (SOR) model, we propose that the stimulus of using RAs (informativeness, product search effectiveness and the lack of sociality stress) can affect consumers' attitude (perceived control and satisfaction), which further affects their behavioural outcomes like impulsive purchase. We validate this research model with survey data from 157 users of RAs. The data largely support the proposed model and indicate that the RAs can significantly contribute to impulsive purchase behaviour in online marketplaces. Theoretical and practical contributions are discussed.

## 1   Introduction

Online marketplaces provide a great deal of product information to consumers. In order to help consumers locate their desired products, all online shopping platforms (such as eBay, Amazon, and TaoBao) have developed various recommendation agents (RAs) to assist consumers in their purchase process. Typical examples of these RAs include iSearch[1], NowDiscount[2] and Taotaosou[3] as well as

---

[1] iSearch is open-source text retrieval software first developed in 1994 by Nassib. iSearch PHP search engine allows to build a searchable database for shopping web site. http://www.isearchthenet.com/isearch/ .
[2] NowDiscount is a shopping assistant for those seeking to find the best local deals, discounts and coupon codes. It includes several advanced features to ensure that you always get the best bang for your buck, including a barcode reader and scanner, price comparison tools, online price match and much more http://nowdiscount.com/nowdiscount_web/.
[3] Taotaosou is the largest images-based searching and shopping engine in China. It covers about 1 million merchants and more than 50,000,000 product pictures. Just inputing a picture, it can help the consumer quickly find the right product in the mass

Amazon's recommendation mechanism of cross selling. The functionality of these RAs has developed from item recommendation, price listing, customer review, product and merchant comparison to more recent image-based searching, collocation, and community-based interactions. From the practical perspective, these new functions have already influenced how people process information and thus play an instrumental role in consumers' decision-making process when buying from online marketplaces.

In the past, research has provided rich insights about the characteristics and the usage of RAs (e.g., Xiao and Benbasat 2007; Hostler, et al. 2011). The impact of RAs as decision support systems on the product selection process has been demonstrated in experimental studies (e.g., Gretzel and Fesenmaier 2006; Wang and Benbasat 2007). In another line of research, an increasing number of studies (e.g., Nass et al. 1994; Zhou et al. 2007) have emphasized that the social perspective could also be applicable to the interactions between human and intelligent technologies, like computers and system software. We argue that an RA is one of the tools in business intelligence that interacts with humans. Therefore, the social perspective can be elaborated and integrated in the new model on the impacts of RAs in online marketplaces.

However, the social aspect of RAs and the effectiveness of RAs in online purchases, especially in motivating consumers' impulsive purchase, has not been verified systematically. In order to address this research gap, the current study aims to provide a conceptual model to explain consumers' online impulsive buying behavior, based on the Stimulus-Organism-Response (SOR) model (Mehrabian and Russell 1974). Applying the tenet of the SOR model in the context of using RAs in online marketplaces, we focus on the impact of using a RA (stimulus) on a consumer's attitudinal response (organism) and the consequent impacts on impulsive purchase behaviour (responses).

Following this introduction, the rest of this paper is structured as below. We develop our conceptual model based on the SOR model and RA-related research in the next section. Section 3 introduces the measures and the survey method. Section 4 explains the data analysis and the results. We conclude the paper with the key findings, implications and future works in Section 5.

## 2 Theoretical Development

### 2.1 Recommendation Agents (RAs)

RAs are commonly regarded as a kind of consumer decision support system (Grenci and Todd 2002). Specifically, RAs can elicit the interests of individual preferences for products and make recommendations accordingly as a shopping assistance. In particular, a RA can learn consumers' shopping preferences when they visit an online store. In addition, when one product is put into the shopping cart, the RA can recommend complementary items for consumers' choice. Some RAs can also compare the attributes of cross-merchant products, especially their prices. They can also provide a product searching function that allows a user to specify certain optimization parameters and runs automatically for the consumer over a period of time (Broeckelmann and Groeppel 2008). As a result, RAs have the potential to support and improve the quality of consumers' decisions making (Haubl and Trifts 2000) by reducing information overload and search complexity (Maes 1994).

As predicted by the theoretical aspect of human information processing, the assistance of RAs affects consumers' shopping behaviour and decision-making processes with respect to the effort exerted during the whole online shopping process (Xiao and Benbasat 2007; Wolfinbarger and Gilly 2001). This subsequently stimulates the change of consumers' attitude and impacts consumers' final purchase behaviour. In the context of using RAs in online shopping, we argue that consumers' effort in decision-making can be significantly reduced by the information processing, alternatives searching, and sociality stress avoidance with the use of a RA.

### 2.2 The Stimulus-Organism-Response (SOR) Model

The SOR model (Mehrabian and Russell 1974) has been widely considered to be the theoretical foundation for research on consumer behaviour. It suggests that stimuli affect and influence a consumer's attitude, responses and emotional states (organism), which consequently result in the customer's behaviour or intention as response. In the context of retail sales, Donovan and Rossiter (1982) were among the first to apply the SOR model in studying consumer behavior. More recently, Floh and Madlberger (2013) used the SOR model to verify the indirect and direct impacts of virtual atmospheric cues on unplanned purchase in the context of e-commerce. Gao and Bai (2014) tested the effect of online travel agents on flow as well as their ultimate influence on consumer satisfaction and purchase intention. In this study, the SOR model provides us with the theoretical basis to investigate

---

product list http://www.taotaosou.com/.

the influence of RA usage on customers' impulsive purchase behaviour. In the context of e-commerce, stimuli can include website design and decorations, internal and external variables, human variables (e.g., Turley and Milliman 2000). Stimuli can leave both direct and indirect impact on attitudinal reaction and behavioural response, such as satisfaction (Gao and Bai 2014), perceived control (Wolfinbarger and Gilly 2001) and impulsive buying behaviour (Floh and Madlberger 2013).

In addition, a recent survey (Floh and Madlberger 2013) suggests that impulsive purchase is unplanned and it is the result of an exposure to stimuli. However, to the best of our knowledge, there is no research taking the SOR model as the reference framework to analyse the relationship between the RA usage and consumers' impulsive purchase behaviour. In the current study, we made use of the SOR model, taking the usage of a RA as the stimulus, a consumer's attitude response as the organism and his/her behaviour outcome as the response. Specifically, Figure 1 shows the conceptual model, aiming to outline the relationships among the usage of a RA (i.e., stimulus including information, searching process and sociality stress), a consumer's attitude response in terms of perceived control and satisfaction as organism, and his/her impulsive purchase as behavioural response. The definitions of the principal constructs are provided in Table 1. We justify each of the proposed hypotheses below.

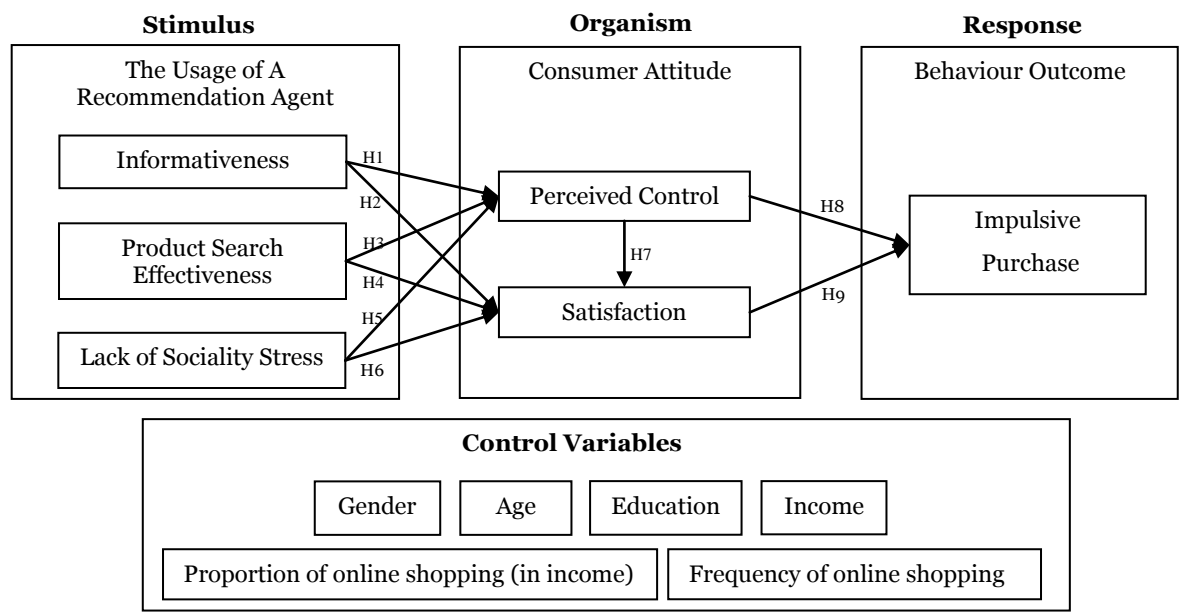

*Figure 1. The Proposed Research Model*

| Constructs | Definitions in This Study | References |
| --- | --- | --- |
| Informativeness | The amount, availability and accuracy of information which a RA provides | Gao and Bai (2014) |
| Product Search Effectiveness | The perceived search effort and time reduced by using a RA | Hostler et al. (2001) |
| Lack of Sociality Stress | The lack of unwanted presence and social interaction from retail sales and the ability to remain anonymous when using a RA | Wolfinbarger and Gilly (2001) |
| Perceived Control | The perceived ability to control a RA during the shopping process | Wolfinbarger and Gilly (2001) |
| Satisfaction | An overall affective evaluation of the experience in using a RA | Carlson and O'Cass (2010) |
| Impulsive Purchase | Purchase which a shopper makes but has not planned in advance | Stern (1962) |

*Table 1. Principal Constructs and Definitions*

## 2.3 The Three Stimuli of Using Recommendation Agents

### 2.3.1 Informativeness

Environmental and individual characteristics are known to influence emotional responses that, in turn, affect behavioural responses. Previous research has shown that RA usage significantly impacts online consumer behaviour (Häubl and Trifts 2000; Lee and Kwon 2008; Hostler et al. 2011). Specifically, informativeness refers to the quantity and quality of information provided by RAs throughout the shopping process (c.f., Gao and Bai 2014). Therefore, if RAs offer useful and meaningful information, consumers can reduce extra effort needed to locate specific information. Informativeness is the most significant factor influencing consumers' decisions with respect to online shopping (Luo, 2002). The online consumer likes to collect and compare rich and useful sources of information when considering the purchase of a product. Besides, the informativeness can help the consumer to evaluate the alternative products and so provide the preference advice to the consumer (Aladwani and Palvia, 2002). High quality information can help the consumer know more about the product and therefore make a better decision, further affecting the attitude toward the shopping website and improving the shopping experience (Elliot and Speck, 2005). Therefore, we argue that RAs can help consumers collect relevant product information and provide a product set of alternatives to reduce the information search time. The information produced in the searching process is one primary external stimulus to consumers to use the RAs during their shopping in online marketplaces.

### 2.3.2 Product Search Effectiveness

To attract the attention of consumers, the website pours a lot of information into consumer's view. However, the overload of information also brings a good deal of trouble to consumers. Due to the consumer's limited processing capacity, informativeness creates a situation of cognitively overloading where consumers receive too much information in a limited time (Malhotra, 1984). If there is no filtering tool to control the quantity and quality of information, consumers become less satisfied, less confident, and more confused. Therefore, we argue that RAs can help consumers to filter and organize the information more efficiently and effectively, allowing them to screen the available alternatives in a personalized way, which refers to the perceived search effort and time reduced by using the RA (c.f., Hostler et al. 2001). Based on a consumer's preference, RAs help screen and sort products as well as provide a comparable list of alternative products, in order to achieve the best match with the consumer's preference. So the product search effectiveness is the other primary external stimulus that may push consumers to use an RA as they shop in online marketplaces.

### 2.3.3 Lack of Sociality Stress

In addition to the searching process and outputs of using RAs, the lack of sociality stress is an important aspect of using an RA. Sociality stress refers to the unwanted presence and social interaction from retail sales people and the inability to remain anonymous when using a RA (c.f. Wolfinbarger and Gilly 2001). This is also considered important in determining consumers' attitude in the related online shopping. In the traditional store shopping environment, the sales persons usually perform the function of product recommendation. In this recommendation process, they cast physical or constant social interactions with consumers. However, it is likely sometimes that the recommendation from a sales person might not be the best choice for a consumer due to the sales person's potentially subjective and incomplete information. Sometimes, the over-enthusiastic attitude of a sales person may also push consumers away.

In the context of e-commerce, RAs are taking over the role of traditional sales persons. Consumers often perceive RAs as both technological mediators and social actors (Qiu and Benbasat 2009). In the same vein, the "Computers are Social Actors" paradigm (Nass et al. 1994) assumes that people will ascribe social aspects when evaluate computer technology. In the case of using online RAs, consumers have complete freedom and control over their own willingness to interact with RAs. Thus, besides the functional design and utilitarian nature of RAs, the potential social influence of RAs, i.e., the lack of social stress in this study, as one of the external stimuli to motivate consumers' purchase, should also be discussed. Accordingly, we propose that the lack of sociality stress of using RAs can be included as a stimulus to a consumer's attitudinal organism.

## 2.4 The Three Stimuli to Attitudinal Organism

When shopping online, consumers may feel lost in the ocean of information. Consumers expect that with the help of RAs, they could get more clear, accurate and useful information. With a strong ability to locate useful, accurate and easy-access information, RAs provide consumers with more sense of control. Following Lee et al. (1990), *perceived control* in this study is defined as perceived situational control, meaning that an individual believes he or she has influence on the event. Most online shopping consumers are prone to have a strong focus of perceived control in the shopping process (Wolfinbarger and Gilly 2001), which can be obtained with the aid of a RA in the shopping process. Thus we propose that:

**Hypothesis1**. *The informativeness of using a RA increases a consumer's perceived control in using the RA during the online shopping process.*

The informativeness offers the access to a diversity of accurate and useful information. The easy access to a variety of useful information can increase consumers' decision-making ability and can enable more perceived control over transaction. Previous research studies have shown that the relevant product information can lead to higher quality decisions (Hostler et al. 2011). The SOR model considers that the stimulus of informativeness has impact on consumer's cognitive states when using RAs, which is the change of their satisfaction. Hausman and Sieke (2009) pointed out that informativeness positively related to the consumer satisfaction. Accordingly, we propose that:

**Hypothesis 2**. *The informativeness of using a RA increases a consumer's satisfaction.*

The effectiveness of the information searching process is one of the essential cues that explains attitudes and is positively related to purchase decisions (Petty et al. 1983; Richard 2005). Moreover, when consumers find the outcomes of the information searching process matches their expectations, they feel cognitively involved which fosters favourable attitudes. Wolfinbarger and Gilly (2001) suggested that consumers anticipate finding selections in online stores where they can have more choices over purchase decisions, for example, to buy special-sized shoes or clothing. Consumers could find it easier to locate suitable products with the help of product searching functions provided by RAs. Significantly, online consumers will perceive more control over the usage of the RA as well as the selections of products. We thus propose that:

**Hypothesis3**. *The product search effectiveness of using a RA increased a consumer's perceived control in using the RA during the online shopping process.*

Product search effectiveness of RAs usage will help consumers focus more on shopping by releasing them from tedious information filtering and facilitating the process of providing preference-related information. Consumers can focus on shopping itself rather than searching for information, and they may be immersed deeply in the purchase process. On the contrary, if consumers consistently encounter useless, limited and unrelated information and are frustrated by the search process, their online shopping experience and attitude could be negatively affected (Zhou et al. 2012). This will result in a decrease of pleasure and satisfaction. We argue that the usage of RAs can improve the product searching effectiveness by recommending product promotion and merchandise information. The product searching effectiveness of RAs usage thus helps consumers to save effort and time spent on information searching and enables satisfactory experiences. We thus propose that:

**Hypothesis4**. *The product search effectiveness of using a RA increases a consumer's satisfaction.*

When shopping in offline stores, consumers may have to interact with sales persons. However, in the online marketplace RAs take the role of sales persons whose main responsibility is to recommend products and to help consumers to make decisions. Through a series of experiments, Nass et al. (1994) suggested that people would apply social relations and expectations to information technologies. Wolfinbarger and Gilly (2001) identified and discussed how the lack of unwanted sociality from retail sales help is one of the attributes that facilitates online shopping. With the absence of physical social interactions with salespeople, crowds and lines, consumers find more freedom and control to complete purchase process. Online consumers do not need to be forced to take all the information from the sales person when using RAs. They do not need to engage in a social interaction while they can focus more on the shopping itself. Therefore, consumers take control over the shopping process and feel unrestrained to find what they need and to make purchase decisions without "having to go through a human being is associated" (Wolfinbarger and Gilly 2001). Moreover, consumers who use RAs can have control over shopping instead of being involved in an unwanted recommendation offered by salespeople. Thus, we propose that:

**Hypothesis5**. *The lack of sociality stress of using a RA increased a consumer's perceived control in using the RA during the online shopping process.*

The lack of sociality stress is appreciated mainly for two reasons. First, sales persons are thought uninformed and subjective, while RAs can provide more comprehensive, objective and accurate information in a great amount. Second, consumers may feel obliged and stressed when interacting with sales persons. Wolfinbarger and Gilly's (2001) experiment proved that sometimes consumers do want to avoid sales persons and lose interest in purchasing when they focus mainly on social interaction and they do not like to feel obligation. Consumers appreciate the less stress of sociality when they purchase online. What's more, the feeling of freedom over shopping may bring them a satisfying shopping experience. Another side of the lack of sociality relates to anonymity. Consumers can avoid the sense of being judged or embarrassed when they try to purchase some private products. Consumers could have more satisfaction when their privacy has been guaranteed. We thus propose that:

***Hypothesis6.*** *The lack of sociality stress of using a RA increases a consumer's satisfaction.*

Research has indicated that perceived control may be the cause of affect (Carver and Scheier 1990). Empirical studies have shown that perceived control affects satisfaction in a diversity of contexts (Langer and Saeger 1977), and specifically shopping satisfaction (Hoffman and Novak 1996). We argue that the more control consumers have during online shopping, the more pleasure consumers may feel. The perceived control may affect satisfaction in a positive way. We thus propose that:

***Hypothesis7.*** *A consumer's perceived control of using a RA increases his/her satisfaction.*

## 2.5 Attitudinal Response to Behaviour Outcome

Several studies have revealed that when involved in the feeling of control, there is a significant possibility that consumers will be prone to overspending (Verplanken and Herabadi 2001). We argue that the improved decision-making efficiency of using RAs can enable consumers to acquire their desired products with unexpected ease during the shopping process. Meanwhile, when consumers feel in control of and know the information of all kinds of products, which are recommended based on their interests and preferences, it is more likely that they will buy those product items that are not originally in their shopping list. We thus propose that:

***Hypothesis8.*** *A consumer's perceived control of using a RA contributes to his/her impulsive purchase behaviour.*

The usage of a RA can influence a consumer's attitude that is directly affected by users' cognitive state and therefore can influence the consumer's behavioural intention and buying behaviour. Zhou et al. (2007) developed a model in which satisfaction was the key construct, referring to a consumer's general feeling about online shopping experience (Bhattacherjee 2001; Delone and McLean 1992). Similarly, Devaraj et al. (2002) suggested that satisfaction was an attitude contributing to a consumer's behavioural intention and subsequently the behaviour. With the satisfactory usage of a RA, it is more likely that a consumer will be more willing to buy the recommended products suggested by the RA, even though these products are not in his/her original buying list. We thus propose that:

***Hypothesis9.*** *A consumer's satisfaction of using a RA contributes to his/her impulsive purchase behaviour.*

The above proposed hypotheses are summarized in Figure 1. We empirically verify the proposed model in the next section.

## 3   Methodology

### 3.1 Measurement development

We adapted the construct measures from the literature into the context of using RAs. The three items to measure informativeness came from Gao and Bai (2014), including the informativeness, knowledgeability and accuracy of the information which a RA provided. The scale for product search effectiveness was based on experimental works by Hostler et al. (2001) as well as Verhagen and van Dolen (2011), covering the perceived effort and time saved by using a RA, and the overall perceived effectiveness of using the RA in the product searching process. Grounded on the study by Kukar-Kinney et al. (2009), we measure the lack of sociality stress by (1) the level of avoiding other shoppers and sales persons; (2) the level of keeping anonymous; as well as (3) the level of receiving useful recommendations even without the interaction with any sales persons when using the RA. Perceived control is measured by two modified items from Wolfinbarger and Gilly (2001) and one additional tailored item for this study, including the levels of feeling no pressure and control over the whole shopping process when using the RA, as well as the evaluation of the control over the interaction with the RA. With respect to satisfaction, we adapt the scale from Hostler et al. (2001) into the context of RA usage, including the evaluation of the RA's design, performance and the level of doing a good job in satisfying the consumer's needs. We made use of the three-item scale from Verhagen and van Dolen (2011) to measure impulsive purchase behaviour. The corresponding measures read "(1) my purchase was unplanned; (2) before using RAs, I did not have the intention to do this shopping; and (3) my actual purchase amount was larger than my planned purchase". For all measures for the principal constructs, we use a 7-point scale with a range between strongly disagree (1) and strongly agree (7) in this study. We also included the demographic data as the control variables for testing the research model.

### 3.2 Data Collection

In this study, we used online questionnaires to collect data. At the beginning of the questionnaire, we first provided a brief background introduction about RAs, indicating that RAs are decision support systems that elicit the interests of individual preferences for products and make recommendations accordingly. In the survey introduction page, we listed a set of popular examples of RAs (i.e. TaoTaoSou, Amazon Product, iSearch, etc.). We then had a filtering question test whether the survey participants had the experience on using a RA. The rest of the survey questions can be answered only by qualified participants based on their most recent RA usage experience.

We collected the data in China. The original English version of the questionnaire was translated first into Chinese and then the Chinese version was translated back into English to examine the reliability of translation. This process was handled by a master student and a PhD student. In order to guarantee the face validity of the questionnaire, we also conducted a pilot data collection with 30 RA users. We used the feedback from this pilot test to revise the questions for better clarity and avoiding any ambiguity. We then sent the final version of the questionnaire to a survey agent who has access to a large pool of online consumers. In order to encourage survey participation, we provided a lucky draw of prizes worth ranging from RMB¥100 to RMB¥3000.

In one month of the sample collection period, 178 questionnaires were collected, among which 157 are valid with complete answers. Table 2 shows the demographic data of these 157 survey respondents, which were later on used in the data analysis.

| Categories | Item | Number | Percentage (%) | Categories | Item | Number | Percentage (%) |
|---|---|---|---|---|---|---|---|
| Gender | Male | 66 | 42.0 | Annual Income (RMB¥) | <30,000 | 37 | 23.6 |
|  | Female | 91 | 58.0 |  | 30,000-60,000 | 48 | 30.6 |
| Age range | <18 | 4 | 2.5 |  | 60,000-96,000 | 52 | 33.1 |
|  | 18-25 | 112 | 71.4 |  | >96,000 | 20 | 12.7 |
|  | 26-30 | 29 | 18.5 | The rate of online shopping | <5% | 32 | 20.4 |
|  | >31 | 12 | 7.6 |  | 5%-10% | 66 | 42.0 |
| Education | High school or below | 6 | 3.8 |  | 11%-30% | 31 | 19.7 |
|  | College/University | 123 | 78.4 |  | 31%-50% | 16 | 10.3 |
|  | Masters or above | 28 | 17.8 |  | >51% | 12 | 7.6 |
| Occupation | Management | 15 | 9.6 | Frequency of online shopping | 3-5 times per week | 26 | 16.6 |
|  | Service | 21 | 13.4 |  | Once a week | 52 | 33.1 |
|  | Sales | 40 | 25.5 |  | Twice a month | 27 | 17.2 |
|  | Maintenance | 11 | 7.0 |  | Once a month | 22 | 14.0 |
|  | Transportation | 8 | 5.1 |  | 3-5 times per year | 18 | 11.5 |
|  | Government | 19 | 12.1 |  | Once a year or less | 12 | 7.6 |
|  | Specialist | 39 | 24.8 |  |  |  |  |
|  | Student | 4 | 2.5 |  |  |  |  |
|  | Others | 66 | 42.0 |  |  |  |  |

*Table 2. Demographic Data of Survey Respondents (n=157)*

## 4   Data Analysis

In this study, we used SmartPLS 2.0 to examine the measurement model and the research model. We tested the confirmatory factor analysis, average variance extracted (AVE), composite reliability, Cronbach's alphas, communality and the correlations between principal constructs. The results documented in Table 3 suggest satisfactory convergent and discriminant validity as well reliability of the measure.  In order to test for multicollinearity, we also conducted collinearity diagnostics. The collinearity indicators (i.e., the tolerance values and variance inflation factors) were all within the acceptable thresholds (Hair et al. 1995).

| Constructs (Item Number) | Composite Reliability | Cronbach's alphas | Communality | 1 | 2 | 3 | 4 | 5 | 6 |
|---|---|---|---|---|---|---|---|---|---|
| 1. Informativeness (3) | 0.926 | 0.880 | 0.806 | **0.90** | | | | | |
| 2. Product search effectiveness (3) | 0.933 | 0.857 | 0.875 | 0.51 | **0.94** | | | | |
| 3. Lack of sociality stress (3) | 0.877 | 0.790 | 0.704 | 0.65 | 0.68 | **0.84** | | | |
| 4. Perceived control (3) | 0.865 | 0.688 | 0.762 | 0.68 | 0.69 | 0.71 | **0.87** | | |
| 5. Satisfaction (3) | 0.931 | 0.888 | 0.818 | 0.80 | 0.75 | 0.69 | 0.72 | **0.90** | |
| 6. Impulsive purchase (3) | 0.899 | 0.834 | 0.748 | 0.57 | 0.51 | 0.53 | 0.49 | 0.62 | **0.87** |

*Table 3. Reliability, Correlation Matrix, and the Square Root of AVE*

After verifying the measurement model, we examined the structural model in SmartPLS 2.0. According to the analysis with the control variables, the proposed research model was largely supported by the survey data, as shown in Figure 2. Specifically, informativeness contributes significantly to perceived control (b=0.28, $0.05<p<0.1$) and satisfaction (b=0.42, $p<0.01$), validating H1 and H2. Product search effectiveness can also significantly increase consumers' perceived control (b=0.20, $0.05<p<0.1$) and satisfaction (b=0.16, $0.05<p<0.1$), verifying H3 and H4. The lack of sociality has a significant positive influence on both perceived control (b=0.33, $p<0.01$) and satisfaction (b=0.21, $0.01<p<0.05$), supporting H5 and H6. The impact of perceived control is significant neither on satisfaction (b=0.14, $p>0.10$) and impulsive purchase (b=0.12, $p>0.10$), thus rejecting H7 and H8. As proposed, satisfaction significantly contributes to impulsive purchase (b=0.41, $p<0.01$), validating H9. None of the control variables of demographic data is significant. The variances explained to the perceived control, satisfaction, and impulsive purchase are 0.542, 0.695 and 0.295 respectively.  The $R^2$ scores for all dependent variables in this study, together with the factor loading, yield an excellent goodness-of-fit for the whole research mode (Chin, 1998).

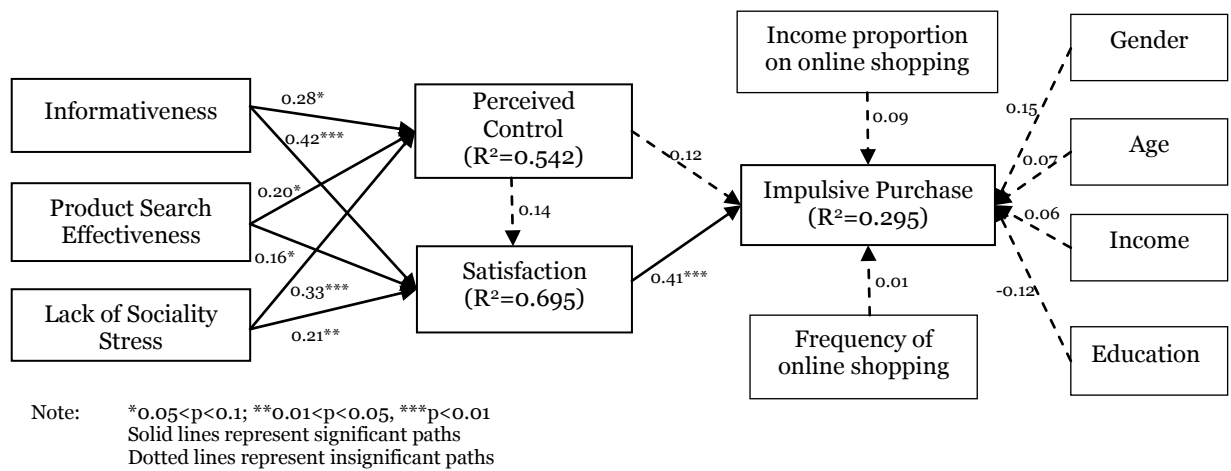

Note:　　*$0.05<p<0.1$; **$0.01<p<0.05$, ***$p<0.01$
　　　　Solid lines represent significant paths
　　　　Dotted lines represent insignificant paths

*Figure 2: SPLS Results with Control Variables*

Given that the data suggested perceived control had a significant effect on neither satisfaction nor impulsive purchase, we conducted additional tests to examine the robustness of perceived control in the model. When taking perceived control as the only independent variable, its causal relationships to satisfaction (b=0.72; $p<0.01$) and to impulsive purchase (b=0.43; $p<0.01$) were both significant. We discussed the implications of these results below.

# 5   Implication and Conclusion

## 5.1 Key Findings and Contribution

In this study, we apply the SOR model to conceptualize the linkages between RAs and impulsive purchases. According to the key findings above, we can conclude that the usage of RAs can significantly affect consumers' attitude change and therefore have impacts on their behaviour outcomes. The current conceptual model extended the SOR model to the context of RA usage and provides a solid theoretical ground for other researchers to further investigate the effects of using RAs in online marketplaces.

One of the key findings in this study is the strong relationship between consumers and the mediated environment, highlighting the positive influence of RAs usage on consumers' decision-making process. According to previous research, the absence of retail workers is appreciated for two reasons: salespeople are often perceived to be unhelpful or uninformed; moreover, they pressure or obligate buyers (Wolfinbarger and Gilly, 2001). Our research suggests that online consumers revel in the fact that they can avoid sales workers online. Participants even sometimes like to avoid helpful salespeople, because they feel obligated to purchase even when they do not really desire to buy an item. Thus, considering transaction online results in shoppers feeling like they do not necessarily have to buy and the lack of people online is associated with freedom. Therefore, when RAs provide interactions and highly personalized recommendation, consumers might foster emotional and social bonds with RAs. Besides the functional design and utilitarian of RAs, the lack of sociality stress of using RAs is conceptualized and empirically confirmed. Our study suggests that consumers appreciate the reduced level of sociality stress when they purchase online. That means the lack of social interactions with salespeople and the anonymity, as well as the feeling of freedom when shopping, can bring consumers significant satisfaction in their shopping experience.

Besides, our data did not support the proposed hypotheses about perceived control. However, the robustness tests verified its important role in contributing to satisfaction, though this direct impact can be overwhelmed by informativeness, product search effectiveness and the lack of sociality stress. These results imply that these three stimuli associated with using the RA are more important and might be more fundamental in developing consumer satisfaction. In addition, the robustness tests verified the full mediation effect of satisfaction between perceived control and impulsive purchase, suggesting the instrumental role of consumer satisfaction in determining their behaviour of impulsive purchase.

## 5.2 Implication

This study has several implications for both researchers and practitioners. Firstly, we attempt to incorporate relevant theories, such as the Stimulus-Organism-Response (SOR) Model of our study. On the basis of our attempts, we can understand more deeply online consumers' shopping behaviour, and also find that the evaluation criteria for information of product and the relationship between the consumer vs. the mediated environment. Our efforts may involve theoretical improvements in explaining the factors of online consumers' impulsive purchase.

Secondly, we systematically evaluated online consumers' impulsive purchase behaviour by examining whether the stimulus of RAs could affect impulsive purchase via perceived control and satisfaction toward websites. In particular, one clue is that time-starved consumers are especially likely to be online shoppers. Another clue is that early and heavy users of the Internet tend to have a strong internal locus of control. These can help both online consumers and online sales to achieve their objectives.

Finally, we clarify that the stimuli associated with RAs involve informativeness, product search effectiveness and lack of sociality stress.  In particular, we determined that three stimuli influence online consumers' purchase intentions. Our findings provide several combinations of successful online shopping sites that perfectly suit online consumers and provide suitable advice to the consumers in the short term. We can provide several combinations of successful web advertisements of online shopping sites that perfectly suit online consumers and help online consumers to locate and be satisfied with what they desire.

## 5.3 Future Research Directions

Otherwise, this study has several limitations and several possible areas for future research. First, this study employed a cross sectional survey method to verify the conceptual model; future research can consider experiments or longitudinal studies, as well as collecting objective data to further verify the proposed model. Second, we only collected the data in one single cultural context. Future research can consider a cross-cultural study to further investigate the potential distinct influences of different environmental stimuli and organisms on behavioural responses in different countries. We look forward to more research that conceptualizes, operationalizes and empirically tests the significance of RAs in online marketplaces.

## Copyright